\documentclass[aps,prl,twocolumn,byrevtex,superscriptaddress]{revtex4-1}

 \usepackage{epsfig}
 \usepackage{color}
 \usepackage{amsmath}
 \usepackage{amsthm}
 \usepackage{amsfonts}
 \usepackage{amssymb}
\usepackage{graphicx}
\usepackage{epstopdf}
\usepackage{ulem}

\def\be{\begin{equation}} \def\eea{\end{eqnarray}}
\def\ee{\end{equation}} \def\bea{\begin{eqnarray}}
\def\ea{\end{array}} \def\ba{\begin{array}}

\newcommand{\bel}[1]{\begin{equation}\label{#1}}

\def\zzz{{\mathchoice {\hbox{$\sf\textstyle Z\kern-0.4em Z$}}
{\hbox{$\sf\scriptstyle Z\kern-0.3em Z$}}
{\hbox{$\sf\scriptscriptstyle Z\kern-0.2em Z$}} {\hbox{$\sf\textstyle
Z\kern-0.4em Z$}}}}

\usepackage{ulem}

\usepackage[dvipsnames]{xcolor}

\begin{document}

\title{Intensity $g^{(2)}$-correlations in random fiber lasers: A random matrix theory approach}

\author{Ernesto~P. Raposo}
\affiliation{Laborat\'orio de F\'{\i}sica Te\'orica e Computacional, Departamento de F\'{\i}sica, Universidade Federal de Pernambuco, 50670-901 Recife, Pernambuco, Brazil}

\author{Iv\'an~R.~R.~Gonz\'alez}
\affiliation{Laborat\'orio de F\'{\i}sica Te\'orica e Computacional, Departamento de F\'{\i}sica, Universidade Federal de Pernambuco, 50670-901 Recife, Pernambuco, Brazil}
\affiliation{Unidade Acadêmica de Belo Jardim, Universidade Federal Rural de Pernambuco, 55156-580 Belo Jardim, Pernambuco, Brazil}
\author{Edwin D. Coronel}
\affiliation{Departamento de F\'{\i}sica, Universidade Federal de Pernambuco, 50670-901 Recife, Pernambuco, Brazil}

\author{Ant\^onio~M.~S. Mac\^edo}
\affiliation{Laborat\'orio de F\'{\i}sica Te\'orica e Computacional, Departamento de F\'{\i}sica, Universidade Federal de Pernambuco, 50670-901 Recife, Pernambuco, Brazil}

\author{Leonardo de S. Menezes}
\affiliation{Chair in Hybrid Nanosystems, Nanoinstitut M\"unchen, Fakult\"at f\"ur Physik, Ludwig-Maximilians-Universit\"at M\"unchen, 80539 M\"unchen, Germany}
\affiliation{Departamento de F\'{\i}sica, Universidade Federal de Pernambuco, 50670-901 Recife, Pernambuco, Brazil}

\author{Raman Kashyap}
\affiliation{Fabulas Laboratory, Department of Engineering Physics, Department of Electrical Engineering, Polytechnique Montreal, Montreal, H3C 3A7, Quebec, Canada}

\author{Anderson S. L. Gomes}
\affiliation{Departamento de F\'{\i}sica, Universidade Federal de Pernambuco, 50670-901 Recife, Pernambuco, Brazil}

\author{Robin Kaiser}
\affiliation{Universit\'e C\^ote d'Azur, CNRS, INPHYNI, 06560 Valbonne, France}



\begin{abstract}
We propose a new approach based on random matrix theory to calculate the temporal second-order intensity correlation function $g^{(2)}(t)$ of the radiation emitted by random lasers and random fiber lasers.
The multimode character of these systems, with a relevant degree of disorder in the active medium, and large number of random scattering centers substantially hinder the calculation of $g^{(2)}(t)$.
Here we apply for the first time in a photonic system the universal statistical properties of Ginibre’s non-Hermitian random matrix ensemble to obtain $g^{(2)}(t)$. 
Excellent agreement is found with time-resolved measurements for several excitation powers of an erbium-based random fiber laser. 
We also discuss the extension of the random matrix approach to address the statistical properties of general disordered photonic systems with various Hamiltonian symmetries.
\end{abstract}
\maketitle


Random lasers (RLs) and random fiber lasers (RFLs) are low-coherence optical sources which have stood out over the last three decades due in part to the ease of fabrication and diverse multidisciplinary applications~\cite{Gomes2021,Turi2014,Baudouin2013}. \linebreak
%
%
Their optical feedback stems from the multiple photon scattering in a disordered active medium~\cite{Letokhov1968}, so differing significantly from the two-mirrors mechanism of the Fabry-Perot type of cavity in conventional lasers.
In particular, RFLs are the quasi-one-dimensional version of RLs~\cite{Matos2007}, employing an optical fiber with embedded gain and randomly distributed scattering centers. 

RFLs have recently attracted a great surge~of interest that led to several new configurations,~much improved experimental characterization, and already important applications
\cite{Gomes2021,Turi2014}.
%
%
However, much less is known, both experimentally and theoretically~\cite{a1,a2,a3,a4}, about~their temporal second-order correlation~function, $g^{(2)}(t)$, a central quantity related to the second order coherence degree, photon statistics, and intensity fluctuations~\cite{gen1}.
%

The theoretical challenge to obtain $g^{(2)}(t)$ for RL and RFL systems is significant due to their unique properties. \linebreak 
The multimode character of RLs and RFLs combined with the intrinsic stochastic dynamics, a relevant disorder degree in the active medium, with many atoms providing the gain, and a large number of random scatterers \linebreak substantially hinder the calculation of $g^{(2)}(t)$ for these systems. 
In this context, standard methods applied~\cite{gen1,gen3,gen4}
to conventional lasers are practically unfeasible. 

In this work, we propose a novel approach to the calculation of the second-order intensity correlation function $g^{(2)}(t)$ in RLs and RFLs based on random matrix theory (RMT)~\cite{rm1,rm3}. 
%
The entries of a random matrix form a set of usually independent random variables, and earlier studies considered statistical ensembles of Gaussian Hermitian random matrices with either orthogonal, unitary, or symplectic properties~\cite{rm1,rm3}.
%
The seminal work by \linebreak Ginibre~\cite{gin} led to a groundbreaking extension of the random matrix formalism to the non-Hermitian counterpart of such ensembles, thus inaugurating a field that is still quite under development~\cite{rm1,rm3},
with striking complexity, rich mathematical structures, and multiple symmetry classes~\cite{38}.
The diversity of physical systems approached by non-Hermitian RMT have burst since then~\cite{rm1,rm3},
from classical diffusion in random media~\cite{ex2}~to complex-energy gapped topological systems~\cite{38}, 
to name a few. 

Here we provide the first application of non-Hermitian RMT to a photonic system. 
The erbium-based RFL~used in this work has been considered in earlier studies of photonic complex behavior such as glassy phase with Parisi's replica symmetry breaking~\cite{n1}, extreme events and L\'evy statistics~\cite{n2,n3,n3b}, and turbulence-like properties~\cite{n4}. 
This system was the first to comprise a remarkably large set of $\sim$$10^3$ specially designed randomly-distributed phase-error-written fiber Bragg gratings, which act as random scatterers of photons~\cite{Gagne2009}. 
Trivalent Er$^{3+}$ erbium ions randomly distributed in the fiber provide the gain that generates the feedback for random lasing emission above the RFL threshold.
%
%
Above threshold a large number $(N \approx 200)$ of longitudinal modes interact, spatially overlap, and stochastically compete for gain~\cite{n1}.

The above mentioned theoretical difficulties are circumvented in the statistical approach of RMT. 
By combining the semiclassical stochastic dynamics driven by the non-Hermitian Hamiltonian of the erbium-based RFL with the universal statistical properties of Ginibre's \linebreak non-Hermitian Gaussian random matrix ensemble with random complex entries (usually termed GinUE~\cite{rm3}),
we obtain an expression for $g^{(2)}(t)$ that compares nicely with time-resolved semiclassical measurements for several excitation powers above threshold in this system. 

The new approach may also prove insightful to deepen the understanding of further statistical properties of RLs, RFLs, and other disordered photonic systems, being thus potentially relevant for advancing as well on elusive issues hard to investigate by conventional methods.
Actually, since our approach is not limited to random lasers (e.g.,~\cite{resub}), this work may have an impact on laser theory in
general. 
Our work may be also able to provide further physical insight on the glassy phase of light in random lasers~\cite{n1}.
%

%
To perform a semiclassical calculation of $g^{(2)}(t)$ for the multimode erbium-based RFL, we start by writing the quantum effective Hamiltonian in the form $\hat{{\cal H}}_{{\scriptsize \mbox{eff}}} = \hat{{\cal H}} +  [ \sum_\lambda i ( \xi_\lambda  + S e^{-i \omega_{{\tiny \mbox{p}}} t} ) \alpha_\lambda^\dagger + h.c.]$.
The field operators associated with the external bath have been formally integrated in $\hat{{\cal H}}$ and the operators of the active transition of the Er$^{3+}$~ions in a two-level model have been treated perturbatively~\cite{hack1,hack3}.
%
So $\hat{{\cal H}}$~is written in the subspace of operators $\alpha_\lambda^\dagger (t)$ and~$\alpha_\lambda(t)$ of creation and annihilation of the system's internal modes~$\lambda$.
$\xi_\lambda(t)$ is a correlated~(nonwhite) quantum noise~\cite{hack1,hack3} that arises from the interaction with the external bath, and $S$ and $\omega_{{\tiny \mbox{p}}}$ are, respectively, the pump amplitude and pump laser central frequency. 
A frame rotating with~$\omega_{{\tiny \mbox{p}}}$~is adopted.

The effective couplings between modes in $\hat{{\cal H}}$ (either~linear and nonlinear) are random owing to the disordered distribution of Er$^{3+}$ ions in the active medium, refractive index with random spatial profile related to the multiple random Bragg gratings inscribed in the fiber, and bath coupling.
We note that previous statistical mechanics approaches to multimode RL systems have considered~\cite{ant0,ant2,ant3}
the mode couplings as Gaussian distributed random variables. 
Moreover, $\hat{{\cal H}}$ is non-Hermitian as the RFL is an open system with losses.

We assume here that $\hat{{\cal H}}$ can be assigned to a member of the Ginibre ensemble of non-Hermitian Gaussian random matrices~(GinUE) 
\cite{rm1,rm3},
with random complex eigenvalues displaying level repulsion and universal statistical properties. 
The eigenvalue density in the complex plane reads~\cite{rm1,rm3,gin}
%
\begin{equation}
\rho(\omega) = \frac{\Gamma(N,|\omega|^2/\sigma^2)}{\pi (N-1)!},
%
%
%
%
\label{1a}
\end{equation}
where $N$ is the matrix order, $\sigma^2$ is the average square modulus of the complex matrix elements, and $\Gamma(z,x)$ is the upper incomplete gamma function.
A striking feature of non-Hermitian random matrices is that the probability distribution of nearest-neighbor level spacings~$s$ in the complex plane is universally cubic as $s \to 0$ in all three Ginibre ensembles (with orthogonal, unitary or symplectic properties)~\cite{rm1,rm3},
%
\begin{equation}
P(s) \sim s^3 e^{-9 \pi s^2/(16 \langle s \rangle^2)},
\label{1b}
\end{equation}
with maximum near the mean level spacing $\langle s \rangle$.
This contrasts with their Hermitian counterparts, whose nonuniversal distributions $P(s) \sim s^\beta e^{-\gamma^2 s^2/\langle s \rangle^2}$ indicate level repulsion degree given by the respective Dyson index $\beta = 1, 2, 4$~\cite{rm1,rm3},
where $\gamma = \Gamma((\beta+2)/2)/\Gamma((\beta+1)/2)$ and $\Gamma(z)$ is the gamma function. 
%
%

The fluctuations in the time series of intensities of  optical spectra of the Er$^{3+}$-based RFL can thus be modelled by a stochastic dynamics governed by a non-Hermitian random matrix.
The Heisenberg equation of motion for $\alpha_\lambda$ yields
\begin{equation}
\frac{d \alpha_\lambda}{dt} = i [\hat{{\cal H}}, \alpha_\lambda ] + S + \xi_\lambda.
\label{1}
\end{equation}
In a semiclassical context, operators $\alpha_\lambda^\dagger (t)$ and $\alpha_\lambda(t)$ are replaced in $\hat{{\cal H}}$ by their complex expected values, i.e.,  we now work with the functional~${\cal H}[\alpha_\lambda^*, \alpha_\lambda]$ instead of the operator $\hat{{\cal H}}[\alpha_\lambda^\dagger, \alpha_\lambda]$.
Further, the noise can be made uncorrelated (white) through a proper choice of basis transformation, $a_\nu (t) = \sum_{\lambda} T_{\nu \lambda} \alpha_\lambda (t)$~\cite{ant3}.
The stochastic semiclassical dynamics of $a_\nu(t)$ is then driven \linebreak by a system of coupled equations with uncorrelated (white) noise $\bar{\xi}_\nu = \sum_\lambda T_{\nu \lambda} \xi_\lambda$.

For excitation powers near the threshold in the random lasing regime, higher-order nonlinear terms in ${\cal H}$ are perturbatively negligible if compared to the quadratic term with random couplings $g_{\lambda \lambda'}$,  
\begin{equation}
{\cal H}[\alpha_\lambda^*, \alpha_\lambda] = -\sum_{\lambda \lambda'} g_{\lambda \lambda'} \alpha_{\lambda}^* \alpha_{\lambda'} + {\cal O}[(\alpha_\lambda^* \alpha_\lambda)^2],
\label{2}
\end{equation}
where $g_{\lambda \lambda'}$ displays nonzero off-diagonal elements due to the openness of the RFL.
In this regime the set of coupled differential equations for $a_\nu(t)$ presents stationary solution $a_\nu(t \to \infty)$ satisfying 
\begin{equation}
\sum_{\nu' \not = \nu} \bar{g}_{\nu \nu'} a_{\nu'}(\infty) + \bar{g}_{\nu \nu} a_{\nu}(\infty) + S_\nu = 0,
\end{equation}
with $\bar{g}_{\nu \nu'} = \sum_{\lambda \lambda'} T_{\nu \lambda} g_{\lambda \lambda'} T_{\nu' \lambda'}^{-1}$ and $S_\nu =  \sum_\lambda T_{\nu \lambda} S$. 
The normal modes have the form $\bar{a}_n(t) = \bar{a}_{n}(\infty) + A_n e^{i \omega_n t}$, where $\omega_n \equiv x_n + i y_n$ denotes the random complex eigenvalues of the non-Hermitian matrix~$\bar{g}_{\nu \nu'}$. 

\begin{figure}[htb] 
\centering
\includegraphics[width=0.97\linewidth]{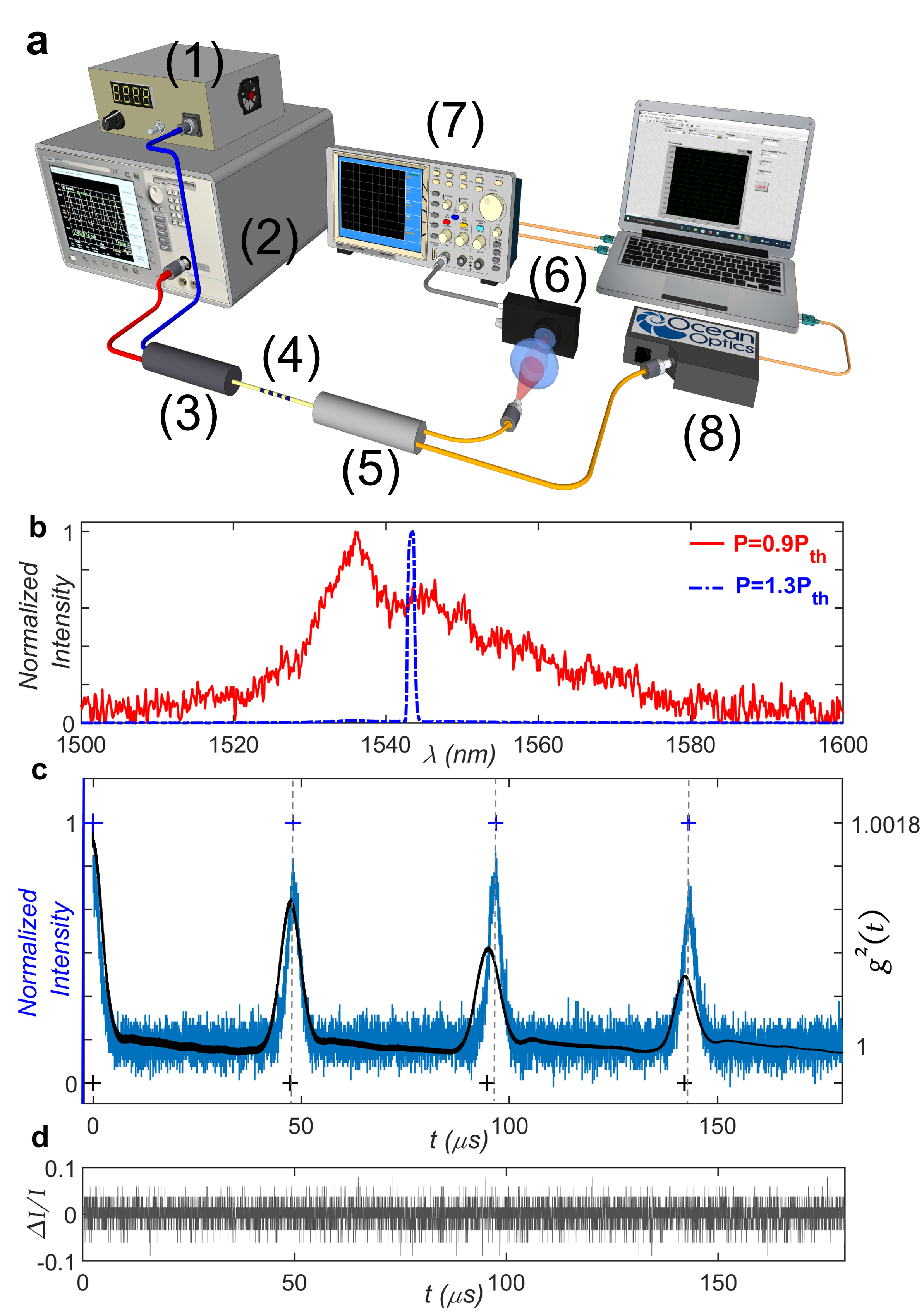}
\caption{(a) Experimental setup of the Er$^{3+}$-based RFL showing (1) the CW pump laser, (2) optical spectrum analyzer (OSA), (3) wavelength division multiplexer (WDM), (4) RFL, (5) coupler, (6) InGaAs photodetector, (7) oscilloscope, and (8) spectrometer.
(b) Spectral profiles below $(P/P_{th} = 0.9)$ and above $(P/P_{th} = 1.3)$ the RFL threshold. 
(c) Intensity signal (blue; light gray in the printed version of the article) and second order~correlation function $g^{(2)}(t)$ (black in both online and printed versions) of the Er$^{3+}$-based RFL~for $P/P_{th} = 4.0$ with effectively Q-switched pulses of nearly same periods, as shown by symbols $+$ at the maxima. 
The intensity was displaced in time so its first maximum coincides with that of $g^{(2)}(t)$.
(d) Homogeneous intensity fluctuations of the pump source.}
\label{fig1}
\end{figure}

In the semiclassical approach the RFL intensity reads $I(t) = \sum_{n} |\bar{a}_n(t)|^2$.
By calculating the temporal second-order intensity correlation function,
\begin{equation}
g^{(2)}(t) = \frac{\langle I(t+t') I(t') \rangle}{\langle I(t') \rangle^2},
\label{6}
\end{equation}
with averages taken over a time interval much larger than the system's relevant time scales, we obtain $g^{(2)}(t)$ in the convenient form,
\begin{equation}
g^{(2)}(t) = 1 + \sum_n b_{n} \cos (x_n t - \varphi_n) e^{-y_n t} 
+ \sum_n c_{n} e^{-2y_nt}, 
\label{7}
\end{equation}
with prefactors $b_n$ and~$c_n$ and phases $\varphi_n$ arising from the modes overlapping integrals at distinct times in~(\ref{6}).
In the GinUE ensemble, eigenvalues and level spacings are distributed as in Eqs.~(\ref{1a}) and~(\ref{1b}), respectively.
This result for $g^{(2)}(t)$ generally applies to RL and RFL systems comprising even quite distinct time scales.
Indeed, the time scales that emerge from the Hamiltonian eigenvalues are naturally set from the fit of the experimental data to Eq.~(\ref{7}) together with Eqs.~(\ref{1a}) and~(\ref{1b}).
For higher excitation powers the addition in Eq.~(\ref{2}) of a fourth-order term with random couplings between a large number of modes leads to a renormalized second-order coupling and Eq.~(\ref{7}) remains approximately valid.

The experimental data of the Er$^{3+}$-based RFL are fitted below to Eq.~(\ref{7}) for several excitation powers. 
Notably, the eigenvalue statistics and universal features of the level repulsion also set the main time scales of~$g^{(2)}(t)$. 

Figure~1(a) shows the experimental setup with a CW home-assembled semiconductor laser operating at 1480~nm as the pump source. 
A polarization-maintaining Er$^{3+}$-doped fiber was used~\cite{Gagne2009,review} (CorActive, absorption peak of 28~dB/m at 1530~nm, numerical aperture 0.25, mode field diameter 5.7~$\mu$m, 30~cm length). 
The RFL output was characterized spectrally using an optical spectrum analyzer (Agilent 86142B) and temporally with a fast (ns time resolution)~InGaAs photodetector (Thorlabs SM05PD5A) and a 300 MHz oscilloscope (Tektronix TDS 3032B).  
The measured RFL threshold~was $P_{th}= 18.7$~mW. 
Figure~1(b) displays the spectral profile below (broadband, relative excitation power $P/P_{th} = 0.9$) and above (narrow band, $P/P_{th} = 1.3$) threshold.

One important experimental aspect is that, above threshold, the RFL operates in an intermittent mode, providing effectively Q-switched intensity pulses whose repetition rate depends on the excitation power~\cite{rate1,rate2}.
Figure~1(c) shows (in blue) an example of such pulses for $P/P_{th} = 4.0$, along with oscillations (in black) of the second-order correlation function~$g^{(2)}(t)$ determined from the intensity signal $I(t)$ using Eq.~(\ref{6}). 
As indicated by the symbols~$+$ that locate the maxima, the oscillation periods of $I(t)$ and~$g^{(2)}(t)$ are nearly the same, with small fluctuations due to the system's stochastic dynamics.

Figure~1(d) presents the intensity fluctuations of the pump source in order to rule out their influence on the pulses profile of Fig.~1(c). 
Moreover, we have additionally determined the relative standard deviation of the pump laser intensity and noticed it remains nearly constant, in contrast with the nonlinear increase of the RFL standard deviation as $P$~is raised above threshold~\cite{n3}.
The stability of the pump source contrasts with the periodic behavior responsible for the photonic Floquet phase recenlty reported in the same system~\cite{floquet}.

\begin{figure}
\centering
\includegraphics[height=5.25 cm,width=0.90\linewidth]{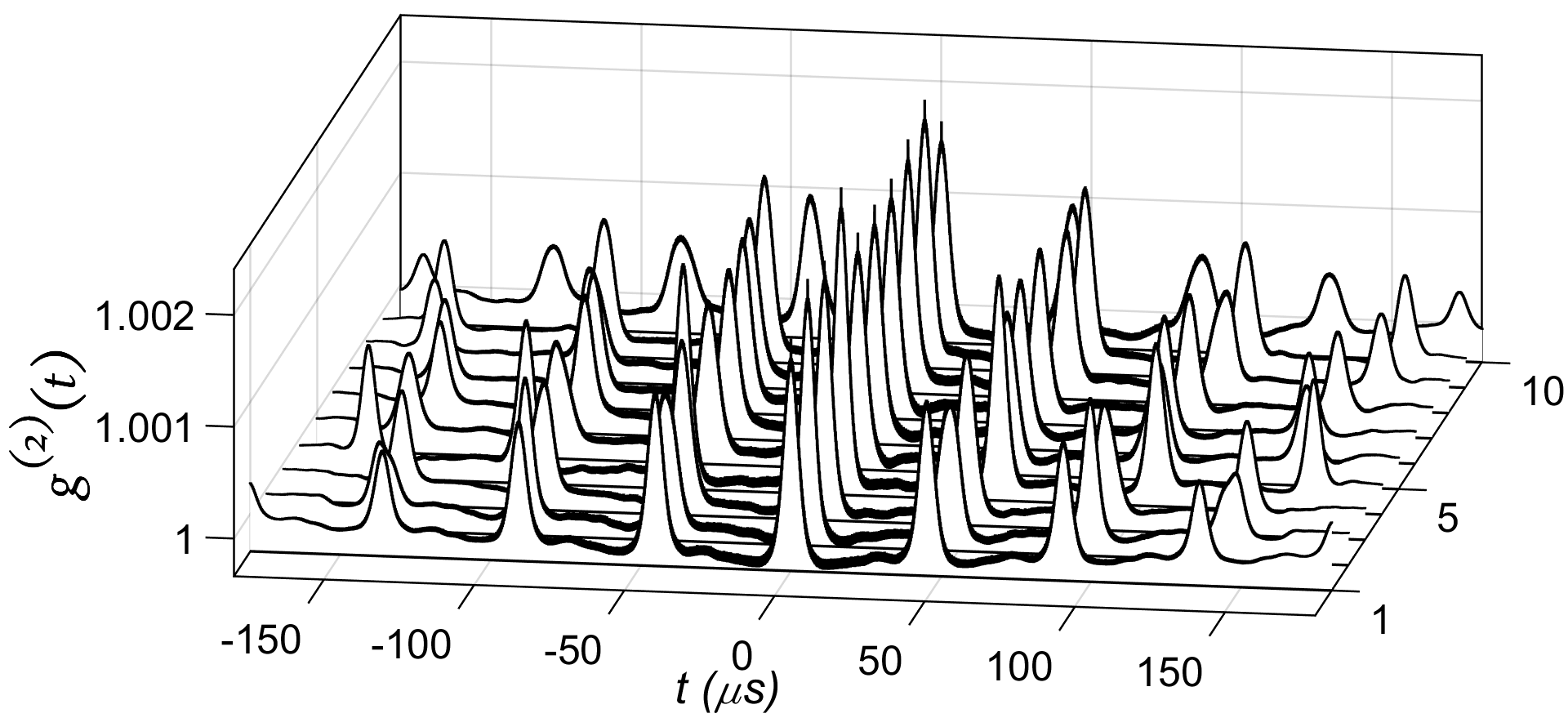}
\caption{Ten measurements of the second order correlation function $g^{(2)}(t)$ of an Er$^{3+}$-based RFL for $P/P_{th} = 4.0$.
Plot-to-plot stochastic variations are noticed leading to slightly distinct model parameters for each curve.}
\label{fig2}
\end{figure}

We show in Fig.~2 ten experimental plots of $g^{(2)}(t)$~of the Er$^{3+}$-based RFL for $P/P_{th} = 4.0$. 
Interestingly, though the overall picture of attenuated oscillations holds for all plots, the results reveal some plot-to-plot variation possibly related to the enhanced eigenvalue sensitivity to perturbations in the random matrix properties of GinUE ensemble, compared to its Hermitian  counterpart~\cite{rm1,rm3}.
%
We denote by $T \sim 50$~$\mu$s the typical separation in time between two consecutive maxima in $g^{(2)}(t)$ observed in Fig.~2 for $P/P_{th} = 4.0$.
A more precise value of~$T$ is obtained below in the context of our model fit to Eq.~(\ref{7}).

The experimental data are nicely described by our theoretical analysis in the random matrix approach. 
We define the nearest-neighbor level spacing in an ordered eigenvalues set by $s_n = (\Delta x_{n}^2 + \Delta y_{n}^2)^{1/2}$, where $\Delta x_{n} = x_{n+1} - x_{n}$ and $\Delta y_{n} = y_{n+1} - y_{n}$.
To generate a sequence of $N$~eigenvalues~$\omega_n$ with level spacing distribution consistent with cubic degree level repulsion, Eq.~(\ref{1b}), we follow a procedure analogous to~\cite{proced}. 
We conveniently introduce the parametrization $\Delta x_{n} = 2 \pi (1 + \delta_n)/T$ and $\Delta y_{n} = \kappa (1 + \gamma_n)$, with $\kappa = 2 \pi [(\langle s \rangle  T/2\pi)^2-1]^{1/2}/T$, $|\delta_n| \ll 1$ and $|\gamma_n| \ll 1$, so that the random spacings~$s_n$ are given by small fluctuations around the mean $\langle s \rangle$.
Also, as the prefactors and phases in Eq.~(\ref{7}) depend on the random eigenvalues $\{ \omega_n \}$, we associate a weight proportional to $\rho(\omega_n)$,  Eq.~(\ref{1a}), with each term in the sums in~(\ref{7}).
In this parametrization $T$~gives the average temporal separation between consecutive maxima in $g^{(2)}(t)$, whereas $\kappa$~governs its long-$t$ envelope exponential decay.
So we set a noteworthy link between the~universal properties of the level spacing and eigenvalue statistics in the random matrix ensemble and the main time scales of the second order correlation function of the~RFL.

\begin{figure}
\centering
\includegraphics[width=1.03\linewidth]{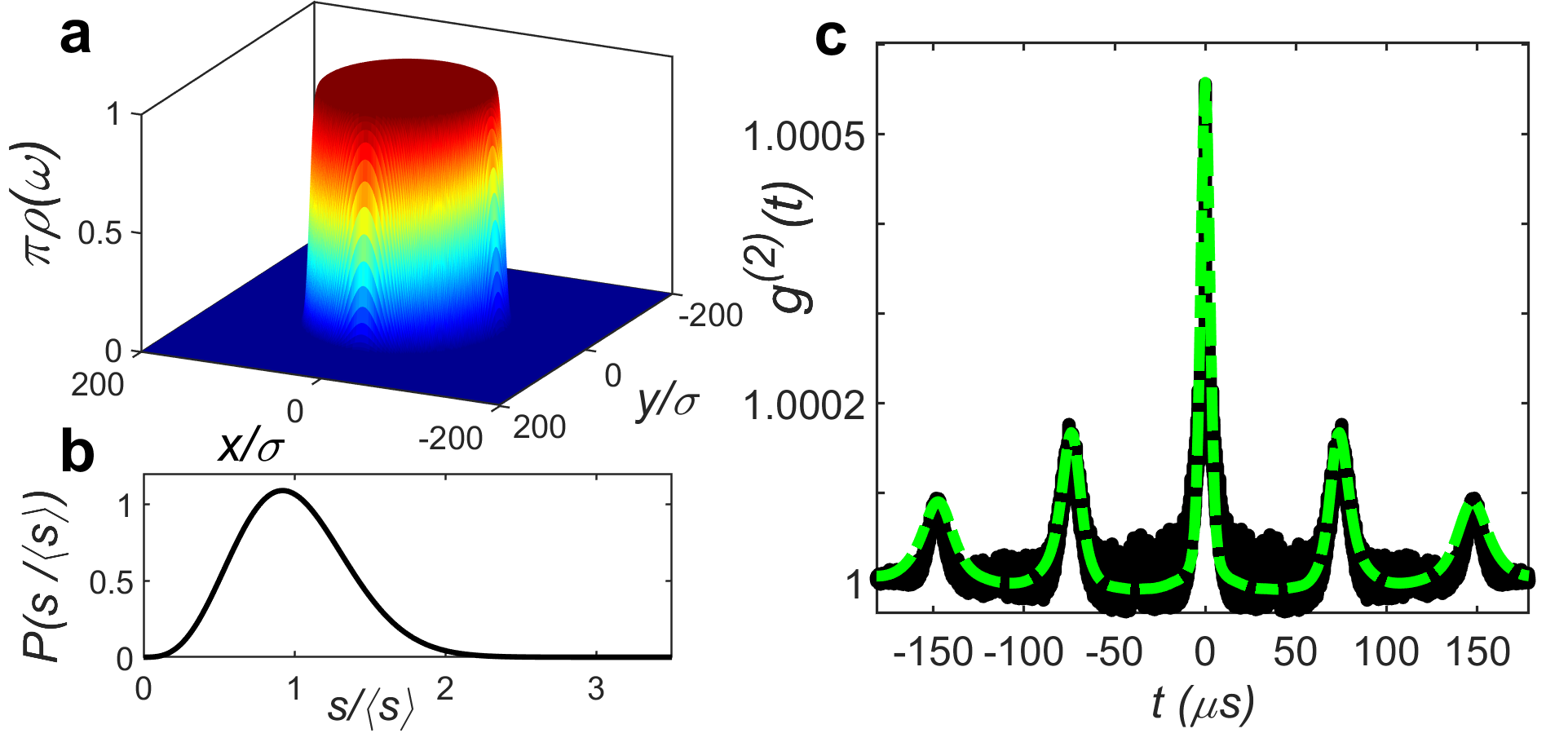}
\caption{(a) Eigenvalue density $\rho(\omega)$, Eq.~(\ref{1a}), in the dimensionless complex plane, with $\omega/\sigma = x/\sigma + iy/\sigma$.
(b)~Probability distribution $P(s)$~of normalized nearest-neighbor level spacings $s/\langle s \rangle$ in the Ginibre ensemble, Eq.~(\ref{1b}). 
(c) One measurement~(black) of the second order correlation function $g^{(2)}(t)$ of the Er$^{3+}$-based RFL near the threshold, $P/P_{th} = 1.6$, and the fit~(dash-dotted green) to the model, Eq.~(\ref{7}). 
Fitting estimates for the oscillation~period and decay constant agree with the RFL~values~(see~text for parameters values).}
\label{fig3}
\end{figure}

%
%
%
Figure~~\ref{fig3}(a) shows  the
eigenvalue density $\rho(\omega)$ in the complex plane, Eq.~(\ref{1a}), using the relation  ${|\omega^2|}/{\sigma^2}=4\pi^2/T^2\sigma^2$, where  $T = 74.0$~$\mu$s and  ${\sigma = 5.88 \times 10^5}$~s$^{-1}$. 
Figure~\ref{fig3}(b) displays the probability distribution $P(s/\langle s \rangle)$ of normalized nearest-neighbor level spacings $s / \langle s \rangle$ in  the Ginibre ensemble, Eq.~(\ref{1b}), with average ${\langle s \rangle = 8.51 \times 10^4}$~s$^{-1}$. 
Figure~\ref{fig3}(c) shows in black circles the experimental data of one measurement of $g^{(2)}(t)$ near the threshold, $P/P_{th} = 1.6$, and the fit to the model result~(\ref{7}) in dash-dotted green lines.  
The model fit value~$T = 74.0$~$\mu$s compares nicely with the experimental measure $T_{\footnotesize{\mbox{exp}}} = 74.3$~$\mu$s.
The remaining fitting parameters are  $b_n = 9.22\times10^{-5}$, $c_n =2.31\times10^{-6}$, and  $\varphi_n = 0.1$ (in order to keep the fitting procedure as simple as possible, we assume that the dependence of $b_n, c_n$, and $\varphi_n$ on the eigenvalues is not too strong and thus effectively work with only one value for each of these three families of parameters). 
We also consider $N = 200$ as the number of~modes of the Er$^{3+}$-based RFL, which was determined using the speckle contrast technique~\cite{n1}. 
These values imply a time constant $\kappa^{-1} \sim 120-300$~$\mu$s consistent with the~lifetime range~\cite{life} of the active state of Er$^{3+}$ ions in the random lasing regime of the CW pumped Er$^{3+}$-based RFL.
Indeed, the observed time scales depend on the Er$^{3+}$ dynamics in the system, so that the typical millisecond Er$^{3+}$ time scale can be in fact lowered to the range of a few hundreds of microseconds when the system operates in the laser regime above threshold, in agreement with our estimates.

Finally, we display in Figs.~4(a) and~4(b) the results for intermediate and high excitation powers, $P/P_{th} = 2.4$ and~$P/P_{th} = 4.0$, respectively. 
As before, a good comparison is found between the periods: $T = 48.0$~$\mu$s and $T_{\footnotesize{\mbox{exp}}} = 48.2$~$\mu$s in Fig.~4(a), while $T = 47.4$~$\mu$s and $T_{\footnotesize{\mbox{exp}}} = 47.5$~$\mu$s in Fig.~4(b).
In particular, the pulse repetition rate $(\propto T^{-1})$ increases monotonically with~$P$, in agreement with results on a random Q-switched fiber~laser~\cite{rate1}.

\begin{figure}
\centering
\includegraphics[width=1.03\linewidth]{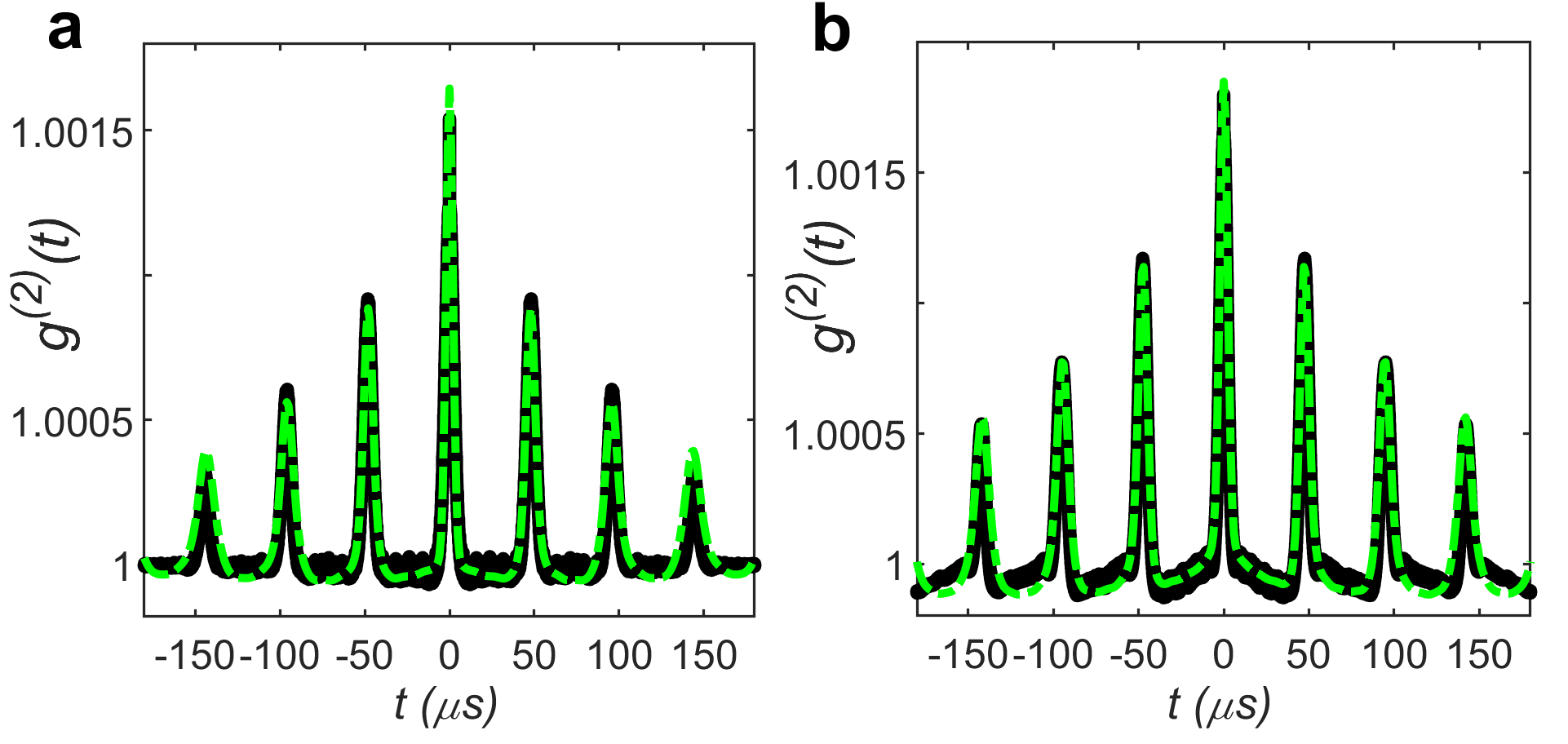}
\caption{Measurements (black) of the second order correlation function $g^{(2)}(t)$ of an Er$^{3+}$-based RFL for (a) intermediate ($P/P_{th} = 2.4$) and (b) high ($P/P_{th} = 4.0$) excitation powers. 
Model results (dash-dotted green), Eq.~(\ref{7}), show nice agreement with the experimental data.}
\label{fig4}
\end{figure}

In conclusion, in this work we have proposed a new approach to the problem of calculating the second-order intensity correlation function in RL and RFL systems, with application to an Er$^{3+}$-based RFL. 
It is difficult to overstate the benefits that the RMT approach can bring to photonic systems exhibiting some kind of disorder.
Rather than working with~a huge set of (virtually unfeasible to determine) disordered mode couplings in the photonic Hamiltonian, the statistical RMT approach takes advantage of the eigenvalue statistics, eigenvector correlators, level spacing density, and repulsion degree, among other features.  

The symmetry properties of each photonic system may guide the proper statistical ensemble to adopt. 
Thus, a diversity of general disordered photonic systems including RLs and RFLs, described by orthogonal, unitary or symplectic Hamiltonian random matrices, either Hermitian or non-Hermitian, with real, complex or quaternionic elements and multiple symmetry classes~\cite{rm1,rm3,gin,38},
can in principle have their statistical emission and further photonic properties addressed by the RMT approach.

We thank the support from CNPq, CAPES, FACEPE, and Instituto Nacional de Fot\^onica (Brazilian agencies).

\bibliographystyle{apsrev4-1}

\end{document}